\documentclass[twocolumn,pdflatex]{aastex631}

\usepackage{here}
\usepackage{natbib}
\usepackage{bm}
\usepackage{mathtools}
\usepackage{amsmath}
\usepackage{amssymb}
\usepackage{mathcomp}
\shorttitle{Radiation Process in Relativistic Alfv\'{e}n Wave}
\shortauthors{Goto \& Asano}

\graphicspath{{./}{figures/}}

\begin{document}

\title{Radiation Process in Relativistic MHD Waves: the Case of Circularly Polarized Alfv\'{e}n Wave}

\author[0000-0003-2825-8903]{Ryota Goto}

\author[0000-0001-9064-160X]{Katsuaki Asano}
\affiliation{Institute for Cosmic Ray Research, The University of Tokyo,
5-1-5 Kashiwanoha, Kashiwa, Chiba 277-8582, Japan}

\begin{abstract}
Turbulence in highly magnetized plasma can be relativistic and induce an electric field comparable to the background magnetic field. Such a strong electric field can affect the emission process of non-thermal electrons. As the first step toward elucidating the emission process in relativistic turbulence, we study the radiation process of electrons in relativistic circularly polarized Alfv\'{e}n waves. While the induced electric field boosts the average energy of low-energy electrons with a Larmor radius smaller than the wavelength, the emissivity for such electrons is suppressed because of the elongated gyro-motion trajectory. The trajectory of high-energy electrons is shaken by the small-scale electric field, which enhances the emissivity. Since the effective Lorentz factor of $E\times B$ drift is $\simeq\sqrt{2}$ in the circularly polarized Alfv\'{e}n waves, the deviation from the standard synchrotron emission is not so prominent. However, a power-law energy injection in the waves can produce a concave photon spectrum, which is similar to the GeV extra component seen in GRB spectra. If the turbulence electric field is responsible for the GeV extra component in GRBs, the estimates of the typical electron energy and magnetic field should be largely altered.

\end{abstract}

\section{Introduction} \label{sec:intro}
The most promising launching mechanism of relativistic jets is the Blandford--Znajek mechanism \citep{1977MNRAS.179..433B}, where the rotation energy of a black hole is extracted via a magnetic field in the ergosphere. In this case, the jet energy is dominated by the magnetic field\citep{2012MNRAS.423.3083M,2014MNRAS.441.3177M}. Relativistic winds from pulsars or magnetars are also dominated by magnetic fields \citep{1969ApJ...157..869G}.

Turbulence may be driven by kink instability or magnetic reconnection in the magnetically dominated (high-$\sigma$) outflows\citep{1998ApJ...493..291B,2009ApJ...700..684M,2011ApJ...728...90M,2014MNRAS.438..278P,2015MNRAS.452.1089P,2016MNRAS.461L..46T,2016MNRAS.456.1739B,2016ApJ...824...48S}.
The turbulence induced in highly magnetized plasma can be responsible for the dissipation of the magnetic field via turbulence reconnection and particle acceleration \citep{2015ApJ...806..167G,2015ApJ...815..101N,2015ApJ...815...16T,2017PhRvL.118e5103Z,2017ApJ...843L..27W,2018MNRAS.481.5687P,2019ApJ...879L..23G,2019ApJ...877...53H,2019ApJ...886..122C,2020ApJ...893L...7W,2020ApJ...895L..40C,2021ApJ...919..111G,2021ApJ...912...48H}. Large amplitude turbulent components of magnetic fields and turbulent motions have been suggested from the observed image and polarization in Crab nebula \citep{2003MNRAS.346..841S,2010MNRAS.405.1809L,2017MNRAS.470.4066B,2023arXiv230916154M}.
In gamma-ray bursts (GRBs) and blazars, the time variability of the flux and polarization may be due to turbulence \citep{2009ApJ...695L..10L,2009MNRAS.394L.117N,2014ApJ...782...92Z,2014ApJ...780...87M}. Thus, the photon emission from relativistic flows in  GRBs, blazars and pulsar wind nebulae (PWNe) can originate from particles in turbulence in high-$\sigma$ plasma.

In high-$\sigma$ plasma, the  Alfv\'{e}n velocity is almost the speed of light. The turbulence velocity can be relativistic. In such turbulence, the induced electric field is comparable to the magnetic field. The electric field affects the trajectory of non-thermal charged particles so that the emission property can be different from the standard synchrotron emission. In most cases, the magnetic fields in the emission region have been estimated by spectral modeling with the standard synchrotron, inverse Compton, and $\pi^0$-decay processes. If the synchrotron emission is largely modified by the turbulence electric field, the estimate of the magnetic field may be misinterpreted. 

In this paper, as the first step toward unveiling the emission property in high-$\sigma$ turbulence, we investigate the radiation process in relativistic circularly polarized Alfv\'{e}n waves, which is analytically described.

The structure of this paper is as follows. In Section \ref{sec:eb}, we review the radiation process in the uniform electric field case. In Section \ref{sec:fluct}, we discuss the emission properties in a relativistic circularly polarized Alfv\'{e}n wave. In Section \ref{sec:radele}, our numerical method to calculate radiation from electron trajectories is shown.
In Section \ref{sec:result1}, we show the numerical results of the radiation for mono-energetic and power-law injections of electrons. In Section \ref{sec:sum}, we summarize our results and discuss their implication for spectral modeling.

\section{Uniform electric field} \label{sec:eb}
In the ideal magnetohydrodynamics (MHD), relativistic turbulece with a turbulent velocity $\delta V\sim c$ induces the motional electric field with a strength $|\bm{E}|=|\frac{\bm{\delta V}}{c}\times \bm{B}|$, which can be comparable to the magnetic field strength $|\bm{B}|$. 
The radiation process of charged particles in such relativistic turbulence can be different from conventional synchrotron radiation. 
In this section, to clarify the effect of an electric field, we first review the radiation process of electrons in the uniform field case.

When a uniform electric field $\bm{E}$ is perpendicular to a uniform magnetic field $\bm{B}$, an observer moving with the drift velocity $\bm{v}_{E\times B}=c\frac{\bm{E}\times \bm{B}}{B^{2}}$ observes zero electric field as long as $E<B$.
The corresponding $E\times B$ drift Lorentz factor is
\begin{eqnarray}
\label{eq:gecb}
\Gamma_{E\times B}=\frac{1}{\sqrt{1-\frac{E^{2}}{B^{2}}}}.
\end{eqnarray}
In this frame (drifting frame), the motion of electrons is spiral around the magnetic field $B'=B/\Gamma_{E\times B}$,
and the emission from electrons is the usual synchrotron radiation in the manetic field $B'$.
(Hereafter, the prime $'$ denotes quantities in the drifting frame.)
Considering electrons isotropically injected with a Lorentz factor $\gamma_{\mathrm{i}}$ in the original frame,
the synchrotron power, namely photon energy emitted per unit time is given by\citep{1979rpa..book.....R,1999AmJPh..67..841J}
\begin{eqnarray}
\label{eq:pb}
P'_{\mathrm{syn}}\simeq2\gamma'^{2}c\sigma_{\mathrm{T}}\frac{B'^{2}}{8\pi},
\end{eqnarray}
where $\gamma'\sim\Gamma_{E\times B}\gamma_{\mathrm{i}}$ is the typical Lorentz factor of the electrons in the drifting frame, and $\sigma_{\mathrm{T}}$ is the Thomson scattering cross section.

In the original frame, the time-averaged Lorentz factor of the electrons is
\begin{eqnarray}
\gamma_{\rm ave}\simeq\Gamma_{E\times B}\gamma'\simeq \Gamma_{E\times B}^2 \gamma_{\mathrm{i}}.
\end{eqnarray}
The average energy is boosted from the injection value by the electric field.
The drift motion of a charged particle in fields of $E\simeq B$ is elongated in the direction of the drift velocity. This asymmetric motion stretches the time interval for $\gamma > \gamma_{\rm ave}$.
As explained in Appendix, this effect slightly increases $\gamma_{\rm ave}$ compared to the above estimate.
The radiation power is Lorentz invariant\citep{1979rpa..book.....R,1999AmJPh..67..841J}. Using $\gamma$, the radiation power is written as 
\begin{eqnarray}
\label{eq:pe}
P_{E\times B}=P'_{\mathrm{syn}}\simeq\frac{1}{\Gamma_{E\times B}^{4}}2\gamma_{\rm ave}^{2}c\sigma_{\mathrm{T}}\frac{B^{2}}{8\pi}.
\end{eqnarray}
The radiation power is suppressed by the factor $1/\Gamma_{E\times B}^{4}$ compared to the usual synchrotron formula without an electric field.
In Appendix, we show more details of analytically calculations of the radiation power.

\section{Radiation in Alfv\'{e}n wave} \label{sec:fluct}

In astrophysical MHD turbulence, there may be a frame where the turbulence is globally isotropic, but locally the fluid velocity and the electromagnetic field are anisotropic. For simplicity, we assume that electrons are isotropically injected in this frame. The turbulence may be injected at a large scale,
and cascade into smaller scales, where the wave amplitude of the turbulence is so small ($\delta V \ll c$) that the effect of the electric field is almost negligible. The induced electric field is significant only at the injection scale.
To investigate the effect of the turbulence electric field on the radiation, we consider a wave propagating to a certain direction with a finite wave length $\lambda$.
While various modes of MHD waves as turbulence are possible, the analytical description is possible for circularly polarized Alfv\'{e}n waves even with non-linear amplitude ($\delta V \sim c$). As a first step, here we focus on this simplest case. 

\subsection{Circularly polarized Alfv\'{e}n wave} \label{sec:alfven}

Relativistic perturbation ($\delta V \sim c$) implies non-linear amplitudes of the perturbed fields. Even in linearly polarized Alfv\'{e}n waves, the induced magnetic pressure leads to compression of fluid.
The compressed part of the waves propagates faster and the nonlinear waves steepen \citep{Shikin1969}, so that the analytical description of the non-linear Alfv\'{e}n waves is difficult.
On the other hand, the circularly polarized Alfv\'{e}n wave can be treated with a constant total pressure in any phase of the wave.

The analytical description of the perturbed electromagnetic fields for a relativistic circularly polarized Alfv\'{e}n wave propagating along the background magnetic field ($z$-axis) is given by \citep{1976JPlPh..15..335K}
\begin{eqnarray}
\label{eq:alfven1}
B_{z}&=&B_{0}, \\
\label{eq:alfven2}
\delta E_{z}&=&0, \\
\label{eq:alfven3}
\delta B_{x}&=&-B_{0}\frac{V}{V_{\rm A}}\cos(kz-\omega_{\rm A} t), \\
\label{eq:alfven4}
\delta B_{y}&=&-B_{0}\frac{V}{V_{\rm A}}\sin(kz-\omega_{\rm A} t), \\
\label{eq:alfven5}
\delta E_{x}&=&-B_{0}\frac{V}{c}\sin(kz-\omega_{\rm A} t), \\
\label{eq:alfven6}
\delta E_{y}&=&B_{0}\frac{V}{c}\cos(kz-\omega_{\rm A} t),
\end{eqnarray}
where $B_0$ is the strength of the background magnetic field, $V$ is the fluid velocity in the Alfv\'{e}n wave, and $k=2\pi/\lambda$ is the wavenumber. Given the gas energy density $\varepsilon$ and the gas pressure $p$, the phase speed of the Alfv\'{e}n wave is written as $V_{\rm A}=\omega_{\rm A}/k=cB_{0}/\sqrt{B_{0}^{2}+4\pi(\varepsilon+p)\Gamma^{2}}$, where $\Gamma=1/\sqrt{1-(V/c)^{2}}$. Introducing the magnetization parameter $\sigma={B_{0}^{2}/(4\pi(\varepsilon+p)\Gamma^{2})}$, the phase speed is rewritten as $V_{\rm A}= c\sqrt{\sigma/(\sigma+1)}$.
As shown in equations (\ref{eq:alfven1})-(\ref{eq:alfven6}), for $V\sim V_{\rm A} \sim c$, the induced electric field is comparable to the background magnetic field.

\subsection{Long wavelength limit} \label{sec:sanaly}

In this section, we consider the case in which the Larmor radius of electrons is much smaller than the wavelength,
\begin{eqnarray}
r_{\rm L0}\equiv\frac{\gamma_{\mathrm{i}}m_{e}c^{2}}{eB_{0}}\ll\lambda.
\end{eqnarray}
In this case, the electric and magnetic fields can be approximated as almost uniform ones.
The radiation power of electrons is suppressed as
 $P_{\mathrm{ave}}\simeq\frac{1}{\Gamma_{E\times B}^{4}}2\gamma_{\mathrm{ave}}^{2}c\sigma_{T}\frac{B^{2}}{8\pi}$ as discussed in section \ref{sec:eb}, where
\begin{eqnarray}
B^2=B_0^2+\delta B^2.
\end{eqnarray}

In the circularly polarized Alfv\'{e}n wave with $B_{0}\simeq\delta B\simeq\delta E$,
the $E\times B$ drift velocity is
\begin{eqnarray}
\label{eq:betaeb}
\beta_{E\times B}=\frac{E}{B}=\frac{\delta E}{\sqrt{B_{0}^{2}+\delta B^{2}}}\simeq\frac{1}{\sqrt{2}},
\end{eqnarray}
which implies $\Gamma_{E\times B}\simeq\sqrt{2}$.
Finally, we obtain
\begin{eqnarray}
P_{\mathrm{ave}}\simeq\gamma_{\mathrm{ave}}^{2}c\sigma_{T}\frac{B^{2}}{16\pi}.
\end{eqnarray}

From the Lorentz transformation of the synchrotron frequency in the $E\times B$ rest frame, the typical frequency is 
\begin{eqnarray}
\label{eq:nupeak}
\nu_{\mathrm{typ}}&&\simeq D\frac{3}{4\pi}\gamma'^{2}\frac{eB'}{m_{e}c}\sin\alpha'\nonumber\\&&\simeq\frac{1}{\Gamma_{E\times B}^{2}}\frac{3}{4\pi}\gamma_{\mathrm{ave}}^{2}\frac{eB}{m_{e}c}\simeq\frac{\frac{4}{\pi}}{\Gamma_{E\times B}^{2}}\nu_{0},
\end{eqnarray}
where $D\simeq\Gamma_{E\times B}$ is the Doppler factor, $B'=B/\Gamma_{E\times B}$, $\gamma'\simeq\gamma_{\mathrm{ave}}/\Gamma_{E\times B}$ and
\begin{eqnarray}
\nu_{0} \equiv \frac{3}{16}\gamma_{\mathrm{ave}}^{2}\frac{eB}{m_{e}c},
\end{eqnarray}
is the typical frequency of isotropic synchrotron radiation.
The typical frequency decreases by the factor $\frac{4}{\pi}/\Gamma_{E\times B}^{2}\simeq2/\pi$ compared to the usual synchrotron formula.

\subsection{Short wavelength limit} \label{sec:lanaly}

In the case with $r_{\rm L0}\gg\lambda$, frequent changes of the field directions should be taken into account.
The radiation power of an electron (charge $-e$) is given by the Li\'{e}nard formula\citep{1949PhRv...75.1912S},
\begin{eqnarray}
\label{eq:powfor1}
P=\frac{2e^{2}}{3c^{3}}\gamma^{4}(a_{\perp}^{2}+\gamma^{2}a_{\parallel}^{2}),
\end{eqnarray}
where $a_{\perp}$ and $a_{\parallel}$ are perpendicular and parallel components of acceleration, respectively.

From the equation of motion
\begin{eqnarray}
\label{eq:eom1}
\frac{d\gamma m_{e}\bm{v}}{dt}=-e\bm{E}-e\frac{\bm{v}}{c}\times\bm{B},
\end{eqnarray}
and the energy conservation
\begin{eqnarray}
\label{eq:ener}
\frac{d\gamma m_{e}c^{2}}{dt}=-e\bm{E}\cdot\bm{v},
\end{eqnarray}
we obtain acceleration of an electron as 
\begin{eqnarray}
\label{eq:a}
\bm{a}=-\frac{e}{\gamma m_{e}}\left(\bm{E}-\frac{\bm{v}}{c}\left(\frac{\bm{v}}{c}\cdot\bm{E}\right)+\frac{\bm{v}}{c}\times\bm{B}\right).
\end{eqnarray}
Then, we obtain the parallel component as
\begin{eqnarray}
\label{eq:apa}
a_{\parallel}=\bm{a}\cdot\frac{\bm{v}}{v}=-\frac{e}{\gamma^{3}m_{e}}\frac{\bm{v}}{v}\cdot\bm{E},
\end{eqnarray}
and the perpendicular component as
\begin{eqnarray}
\label{eq:ape}
\bm{a}_{\perp}&=&\bm{a}-a_{\parallel}\frac{\bm{v}}{v}\nonumber\\&=&-\frac{e}{\gamma m_{e}}\left(\bm{E}-\frac{\bm{v}}{v}\left(\frac{\bm{v}}{v}\cdot\bm{E}\right)+\frac{\bm{v}}{c}\times\bm{B}\right).
\end{eqnarray}
For $\gamma\gg1$, equations (\ref{eq:apa}) and (\ref{eq:ape}) imply
\begin{eqnarray}
\label{eq:pape}
a_{\perp}^{2}+\gamma^{2}a_{\parallel}^{2}\simeq a_{\perp}^{2},
\end{eqnarray}
%This approximation is mostly valid because we are considering ideal MHD electric field $E<B$.
which leads to\citep{1949PhRv...75.1912S}
\begin{eqnarray}
\label{eq:powapl}
&&P\simeq \frac{\gamma^{2}c\sigma_{T}}{4\pi}\left( \bm{E}-\frac{\bm{v}}{v}\bigg\lparen\frac{\bm{v}}{v}\cdot\bm{E}\bigg\rparen+\frac{\bm{v}}{c}\times\bm{B} \right)^{2}.
\end{eqnarray}

From equation (\ref{eq:ener}) with $\delta B\simeq\delta E\simeq B_{0}$, the fractional change of the Lorentz factor during the wave crossing time $\Delta t=\lambda/v$ is
\begin{eqnarray}
\label{eq:enerw}
\frac{\Delta\gamma}{\gamma}\sim\frac{\Delta t}{\gamma}\frac{d\gamma}{dt}\sim \frac{\lambda e B_0}{\gamma m_{e}c^{2}}\sim\frac{\lambda}{r_{\rm L0}}\ll1.
\end{eqnarray}
Similarly, equation (\ref{eq:eom1}) gives the angle change as 
\begin{eqnarray}
\label{eq:angle}
\Delta\theta\simeq\frac{\Delta t}{\gamma m_{e}v}\frac{d\gamma m_{e}v_{\perp}}{dt}\simeq \frac{\lambda e B_0}{\gamma m_{e}c^{2}}\simeq\frac{\lambda}{r_{\rm L0}}\ll1.
\end{eqnarray} 

Therefore, the electron trajectory is a spiral motion around the background magnetic field with a small oscillation.
The velocity components of an electron injected with a pitch angle $\theta_{\rm i}$ and an azimuthal angle $\phi_{\rm i}$ are approximately expressed as
\begin{eqnarray}
\label{eq:gl}
\gamma&\simeq&\gamma_{\mathrm{i}}\simeq\gamma_{\mathrm{ave}},
\\
\label{eq:tralx}
v_{x}&\simeq& v\sin\theta_{\rm i}\cos(\omega_{B}t+\phi_{\rm i}),
\\
\label{eq:traly}
v_{y}&\simeq& v\sin\theta_{\rm i}\sin(\omega_{B}t+\phi_{\rm i}),
\\
\label{eq:tralz}
v_{z}&\simeq& v\cos\theta_{\rm i},
\end{eqnarray}
where $\omega_{B}=\frac{eB_{0}}{\gamma m_{e}c}$ is the gyro frequency.

As the electron injection is isotropic in this frame, we average over the angles $\theta_{\rm i}$, $\phi_{\rm i}$, and a time interval $T$ much longer than the wave crossing time $\lambda/v$.
Equations (\ref{eq:alfven1})-(\ref{eq:alfven6}), (\ref{eq:powapl}) and (\ref{eq:gl})-(\ref{eq:tralz}) lead to
\begin{eqnarray}
\label{eq:powavel}
&&P_{\mathrm{ave}}\equiv\frac{1}{4\pi}\int_{0}^{2\pi}d\phi_{\rm i}\int_{0}^{\pi}d\theta_{\rm i}\sin\theta_{\rm i} \frac{1}{T}\int_{0}^{T}dtP(\gamma_{\rm i},\theta_{\rm i},\phi_{\rm i},t)\nonumber\\
&&\simeq\frac{4}{3}\gamma_{\mathrm{ave}}^{2}c\sigma_{T}\bigg\lparen\frac{B_{0}^{2}}{8\pi}+\frac{\delta B^{2}}{8\pi}+\frac{\delta E^{2}}{8\pi}\bigg\rparen
\simeq\gamma_{\mathrm{ave}}^{2}c\sigma_{T}\frac{B^{2}}{4\pi},
\end{eqnarray}
where $B^2=B_0^2+\delta B^2$ again.
Differently from the case for $\lambda\gg r_{\rm L0}$ in section \ref{sec:sanaly}, the radiation power is rather enhanced by the perturbed electric field.

We estimate the typical emission frequency.
As we consider MHD turbulence, the Larmor radius of non-relativistic electrons is assumed to be short enough as $\lambda\gg m_{e}c^{2}/eB$. In this case, equation (\ref{eq:angle}) implies $\lambda \gg r_{\rm L,0}/\gamma$, and $\Delta\theta \gg 1/\gamma$.
This means that the radiation cone with opening angle $1/\gamma$ sweeps a certain position before an electron oscillates with a deflection angle $\Delta\theta$\citep{2000ApJ...540..704M}.
With $a \simeq a_{\perp}$, the emission frequency can be estimated \citep{2010ApJ...724.1283R} as
\begin{eqnarray}
\label{eq:nu}
\nu(\theta_{\rm i},\phi_{\rm i},t)\equiv\frac{3\gamma^{3}(\theta_{\rm i},\phi_{\rm i},t)a_{\perp}(\theta_{\rm i},\phi_{\rm i},t)}{4\pi c}.
\end{eqnarray}

Averaging with the weight of the radiation power over the angles $\theta_{\rm i},\phi_{\rm i}$ and time, we can obtain the typical emitted frequency
\begin{eqnarray}
\label{eq:nutyps}
&&\nu_{\mathrm{typ}}\equiv\frac{1}{{P_{\mathrm{ave}}}}\frac{1}{4\pi}\int_{0}^{2\pi}d\phi_{\rm i}\int_{0}^{\pi}d\theta_{\rm i}\sin\theta_{\rm i}\nonumber\\&&\times \frac{1}{T}\int_{0}^{T}dt\nu(\theta_{\rm i},\phi_{\rm i},t)P(\theta_{\rm i},\phi_{\rm i},t).
\end{eqnarray}
In the case without the turbulence, assuming $\delta E=\delta B=0$ and using equations (\ref{eq:alfven1})-(\ref{eq:alfven6}), (\ref{eq:ape}), (\ref{eq:powapl}) and (\ref{eq:gl})-(\ref{eq:tralz}), we numerically obtain
\begin{eqnarray}
\label{eq:nutyps}
&&\nu_{\mathrm{typ}}\simeq 1.46\nu_{0}.
\end{eqnarray}

In the circularly polarized Alfv\'{e}n wave, a similar calculation assuming $\delta E\simeq\delta B\simeq B_{0}$ leads to
\begin{eqnarray}
\label{eq:nutyps}
\nu_{\mathrm{typ}}\simeq2.11\nu_{0}.
\end{eqnarray}

The increase of the peak frequency is because the perturbed electric field $\delta E$ increases the perpendicular acceleration $a_{\perp}$ in equation (\ref{eq:ape}).

\section{Numerical Method}
\label{sec:radele}

As we have mentioned, we assume that a monochromatic wave propagates along the $z$-axis as expressed by equations (\ref{eq:alfven1})-(\ref{eq:alfven6}). The wave locally dominates the turbulence, which is globally isotropic. In the wave rest frame (moving along the $z$-axis with the velocity $V_{\rm A}$), the electric field vanishes. Neglecting the radiative cooling, the electron energy in the wave rest frame is conserved. The electron energy in the original frame periodically oscillates following the change of momentum in the wave rest frame. Therefore, the time averaged Lorentz factor in the original frame does not evolve in this circularly polarized Alfv\'{e}n wave.

We isotropically inject electrons in the original frame, and solve the equations of motion
\begin{eqnarray}
\label{eq:position}
\frac{d\bm{x}}{dt}&=&\bm{v}, \\
\label{eq:eom}
\frac{d\gamma m_{e}\bm{v}}{dt}&=&-e\bm{E}(\bm{x},t)-e\frac{\bm{v}}{c}\times\bm{B}(\bm{x},t),
\end{eqnarray}
using the Boris-C solver with second-order accuracy in \citet{2018PhPl...25k2110Z}
with a time step $\Delta t=0.02\min(\lambda/c,r_{\rm L0}/c)$.
We isotropically inject 1600 electrons in total.

The average Lorentz factor for electrons injected with a Lorentz factor $\gamma_{\mathrm{i}}\gg1$ is calculated with a time interval $T=200\min(\lambda/c,r_{\rm L0}/c)$ as
\begin{eqnarray}
\label{eq:gave}
\gamma_{\mathrm{ave}}(\gamma_{\mathrm{i}})\equiv&&\frac{1}{4\pi}\int_{0}^{2\pi}d\phi_{\rm i}\int_{0}^{\pi}d\theta_{\rm i}\sin\theta_{\rm i}\frac{1}{T}\int_{0}^{T}dt\gamma.
\end{eqnarray}

The radiation power is numerically calculated using the Li\'{e}nard formula
\begin{eqnarray}
\label{eq:powfor}
P(\gamma_{\mathrm{i}},\theta_{\rm i},\phi_{\rm i},t)=\frac{2e^{2}}{3c^{3}}\gamma^{4}(a_{\perp}^{2}+\gamma^{2}a_{\parallel}^{2}),
\end{eqnarray}
which is also averaged as
\begin{eqnarray}
\label{eq:powave}
P_{\mathrm{ave}}(\gamma_{\mathrm{ave}}(\gamma_{\mathrm{i}}))\equiv&&\frac{1}{4\pi}\int_{0}^{2\pi}d\phi_{\rm i}\int_{0}^{\pi}d\theta_{\rm i}\sin\theta_{\rm i}\nonumber\\&&\times\frac{1}{T}\int_{0}^{T}dtP(\gamma_{\rm i},\theta_{\rm i},\phi_{\rm i},t).
\end{eqnarray}

The radiation power spectrum of an electron is given by the Fourier transformation of the radiation electric field described by the Li\'{e}nard Wiechart potential \citep{1979rpa..book.....R,1999AmJPh..67..841J}.
For $\lambda\gg m_{e}c^{2}/eB$ and $E\leq B$, the radiation spectrum can be approximately calculated \citep{2010ApJ...724.1283R} by
\begin{eqnarray}
P(\nu,\gamma_{\mathrm{i}},\theta_{\mathrm{i}},\phi_{\mathrm{i}})=\frac{1}{T}\int_{0}^{T}dt\frac{\sqrt{3}e^{2}\gamma a_{\perp}}{c^{2}}F\bigg(\frac{4\pi c\nu}{3\gamma^{3}a_{\perp}}\bigg),
\end{eqnarray}
where $F(x) \equiv x\int_{x}^{\infty}K_{\frac{5}{3}}(\xi)d\xi$, and $K_{\frac{5}{3}}(\xi)$ is the modified Bessel function of the order $5/3$. We resolve frequency range by 20 meshes per logbin. For calculation of $F(x)$, we interpolate the analytical approximate formula in \citet{2013RAA....13..680F} by the cubic spline in every time step $\Delta t$. 

The radiation spectrum is averaged over the injection angles $\theta_{\rm i}$ and $\phi_{\rm i}$ as
\begin{eqnarray}
\label{eq:curspectrumave}
P_{\nu,{\rm ave}}(\gamma_{\mathrm{ave}}(\gamma_{\mathrm{i}}))
=&&\frac{1}{4\pi}\int_{0}^{2\pi}d\phi_{\mathrm{i}}\int_{0}^{\pi}d\theta_{\mathrm{i}}\sin\theta_{\mathrm{i}}\nonumber\\&&\times P(\nu,\gamma_{\mathrm{i}},\theta_{\mathrm{i}},\phi_{\mathrm{i}}).
\end{eqnarray}

\section{Mono-Energetic Injection}
\label{sec:result1}

In this section, we show numerical results for electrons isotropically injected with an initial Lorentz factor $\gamma_{\mathrm{i}}$ for a parameter range of $10^{-5}\leq r_{\rm L0}/\lambda\leq10^{3}$.
As we mentioned, the MHD approximation implies $\lambda\gg m_{e}c^{2}/eB$.
So the jitter radiation mechanism \citep[$\lambda < m_{e}c^{2}/eB$,][]{2000ApJ...540..704M} is not eligible in our case.
Our results are shown normalized by the well-known formulae for a uniform magnetic field: the radiation power
\begin{eqnarray}
P_0 \equiv \frac{4}{3}\gamma_{\mathrm{ave}}^{2}c\sigma_{T}\frac{B^{2}}{8\pi},
\end{eqnarray}
and the typical frequency $\nu_{0}$ given by eq. (\ref{eq:nupeak}).

\begin{figure}[ht]
  \centering
  \includegraphics[width=9cm]{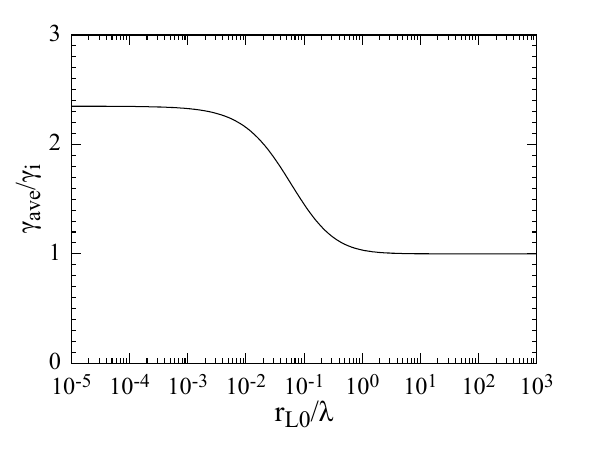}
  \includegraphics[width=9cm]{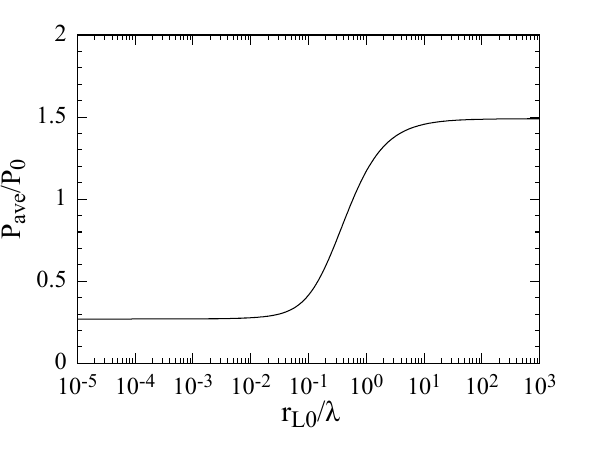}
  \caption{Upper: the average Lorentz factor $\gamma_{\mathrm{ave}}$ of electrons injected with Lorentz factor $\gamma_{\mathrm{i}}$. Lower: the average radiation power $P_{\mathrm{ave}}$ normalized by $P_0$.}
  \label{fig:gpowaverl}
\end{figure}

The upper panel of Figure \ref{fig:gpowaverl} shows the average Lorentz factor $\gamma_{\mathrm{ave}}$ as a function of $r_{\rm L0}/\lambda$.
In low energy limit ($r_{\rm L0}/\lambda\ll1$), the electron energy is boosted by a factor of $2\times1.19\simeq 2.4$ as explained in section \ref{sec:eb} and Appendix, while $\gamma_{\mathrm{ave}}$ is almost the sama as $\gamma_{\mathrm{i}}$ in high energy limit ($r_{\rm L0}/\lambda\ll1$) as explained in \S \ref{sec:lanaly}.
As shown in the lower panel of Figure \ref{fig:gpowaverl}, 
the radiation power for $r_{\rm L0}/\lambda\ll1$ is suppressed by a factor of $1/\Gamma_{E\times B}^{4}\simeq1/4$ as estimated in equation (\ref{eq:powavegaves}).
On the other hand, the power is $1.5$-fold enhanced by the electric field for $r_{\rm L0}/\lambda\gg1$ as discussed in
\S \ref{sec:lanaly}.

\begin{figure}[ht]
  \centering
  \includegraphics[width=8.7cm]{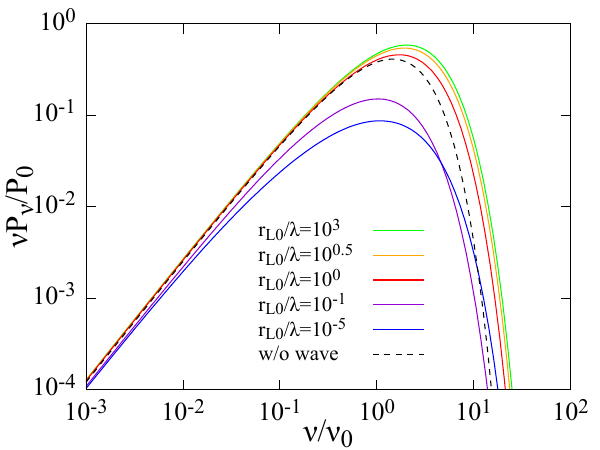}
  \caption{Radiation spectra for different  $r_{\rm L0}/\lambda$. The synchrotron spectrum without the turbulence is shown with black line as a reference.} 
  \label{fig:sperl}
\end{figure}

In Figure \ref{fig:sperl}, we show the radiation spectra $\nu P_{\nu,\rm ave}/P_{0}$.
In the case of $r_{\rm L0}/\lambda=10^{3}$ (green), the radiation power is enhanced compared to the case without the turbulence (black).  
The peak frequency is also increased compared to the synchrotron case
as discussed in \S \ref{sec:lanaly}.

For $r_{\rm L0}/\lambda=10^{-5}$ (blue), the radiation power is suppressed compared to the case without the turbulence by a factor of $1/\Gamma_{E\times B}\simeq1/4$.
The slight decrease in the typical frequency is consistent with the discussion in equation (\ref{eq:nupeak}).

The spectrum for $r_{\rm L0}/\lambda=10^{-5}$ is broader than the other cases.
The broadening is due to the dispersions in the Doppler factor $D=\Gamma_{E\times B}(1+\beta_{E\times B}\cos\theta')$ and the Lorentz factor in the $E\times B$ rest frame $\gamma'=\Gamma_{E\times B}\gamma_{\mathrm{i}}(1-\beta_{E\times B}\beta_{0}\cos\theta)$, where $\theta'$ and $\theta$ are the angles of the propagation direction of photons and the electron motion direction at injection, respectively, with respect to the $E\times B$ drift velocity. The dispersions are significant for $r_{\rm L0} \ll \lambda$.

\section{Power-Law Injection} \label{sec:raddis}

In a strong magnetic field, the effect of radiative cooling appears in the high-energy electron energy distribution. Assuming a continuous electron injection with a power-law energy distribution and electron escape, a broken power-law energy distribution is frequently assumed as a steady state.
In our case with relativistic turbulence, we need to consider not only cooling but also the energy boost by the electric field, shown in Figure \ref{fig:gpowaverl}.

\subsection{Method} \label{sec:method2}

Depending on the cooling break in the electron energy distribution, the photon spectrum shows a variety. We numerically calculate the time evolution of electron energy distribution $N(\gamma,t)$ in a relativistic Alfv\'{e}n wave by solving the equation of continuity in the energy space as
\begin{eqnarray}
\label{eq:con}
&&\frac{\partial N(\gamma_{\mathrm{ave}},t)}{\partial t}+\frac{\partial}{\partial\gamma_{\mathrm{ave}}}(\dot\gamma_{\mathrm{ave}}(\gamma_{\mathrm{ave}}) N(\gamma_{\mathrm{ave}},t))\nonumber\\&&=\dot N_{\mathrm{inj}}(\gamma_{\mathrm{ave}}),
\end{eqnarray}
where $\dot\gamma_{\mathrm{ave}}(\gamma_{\mathrm{ave}})$ is radiative cooling rate of electrons in relativistic Alfv\'{e}n wave and the injection term $\dot N_{\mathrm{inj}}(\gamma_{\mathrm{ave}})$ is assumed to be constant with time. By changing the calculation time $t$, the energy at the cooling break can be adjusted.

As shown in the upper panel of Figure \ref{fig:gpowaverl}, we take into account the boost of Lorentz factor of electrons after injection for the injected electron energy distribution.
The initial electron energy distribution at injection is assumed as a power law distribution with an exponential cutoff as 
\begin{eqnarray}
\label{eq:ninj0}
\dot N_{\mathrm{inj},0}(\gamma_{\mathrm{i}})&=&C\gamma_{\mathrm{i}}^{-p}\exp\bigg\lparen-\frac{\gamma_{\mathrm{i}}}{\gamma_{\mathrm{cut}}}\bigg\rparen\nonumber\\&&\mathrm{for}\ \ \ \gamma_{\mathrm{min}}<\gamma_{\mathrm{i}}<\gamma_{\mathrm{max}},
\end{eqnarray}
where $\gamma_{\mathrm{i}}$ is the initial Lorentz factor at injection and  $C=1/\int_{\gamma_{\mathrm{min}}}^{\gamma_{\mathrm{max}}}d\gamma_{\mathrm{i}}\dot N_{\mathrm{inj},0}(\gamma_{\mathrm{i}})t$ is the normalization factor.
In this paper, we adopt $p=2$.
Then, we calculate the electron distribution after the energy boost by the wave using $\gamma_{\mathrm{ave}}(\gamma_{\mathrm{i}})$ calculated with equation (\ref{eq:gave}) as
\begin{eqnarray}
\label{eq:ninj}
\dot N_{\mathrm{inj}}(\gamma_{\mathrm{ave}})=\frac{d\gamma_{\mathrm{i}}}{d\gamma_{\mathrm{ave}}}\dot N_{\mathrm{inj},0}(\gamma_{\mathrm{i}}(\gamma_{\mathrm{ave}})).
\end{eqnarray}

For the radiative cooling rate $\dot\gamma_{\mathrm{ave}}(\gamma_{\mathrm{ave}})$, we use the average radiation power $P_{\mathrm{ave}}(\gamma_{\mathrm{ave}})$ calculated by equation (\ref{eq:powave}) as
\begin{eqnarray}
\label{eq:gdot}
\dot\gamma_{\mathrm{ave}}(\gamma_{\mathrm{ave}})=-\frac{P_{\mathrm{ave}}(\gamma_{\mathrm{ave}})}{m_{e}c^{2}}.
\end{eqnarray}

The radiation spectrum is calculated as
\begin{eqnarray}
\label{eq:powdis}
P_{\nu}(t)=\int_{1}^{\infty}d\gamma_{\mathrm{ave}} N(\gamma_{\mathrm{ave}},t)P_{\nu,\mathrm{ave}}(\gamma_{\mathrm{ave}}),
\end{eqnarray}
where $P_{\nu,{\rm ave}}(\gamma_{\mathrm{ave}})$ is the radiation spectrum calculated with eq. (\ref{eq:curspectrumave}).

\subsection{Slow cooling} \label{sec:result2}
The cooling break should appear at $\gamma_{\mathrm{ave}}=\gamma_{\mathrm{c}}$, which satisfies $t=\gamma_{\mathrm{ave}}/\dot{\gamma}_{\mathrm{ave}}$.
First, we consider the case with $\gamma_{\mathrm{c}}\gg \gamma_{\mathrm{cut}}$, where the cooling effect is negligible (slow cooling).
As we have discussed, the emission properties change at $\gamma_{\mathrm{ave}}=\gamma_\lambda$, where $\gamma_\lambda$ is the electron Lorentz factor at which the corresponding Larmor radius is comparable to the wavelength,
\begin{eqnarray}
\label{eq:glambda}
\frac{\gamma_{\lambda}m_{e}c^{2}}{eB_{0}}=\lambda.
\end{eqnarray}

\begin{figure}[ht]
  \centering
  \includegraphics[width=9cm]{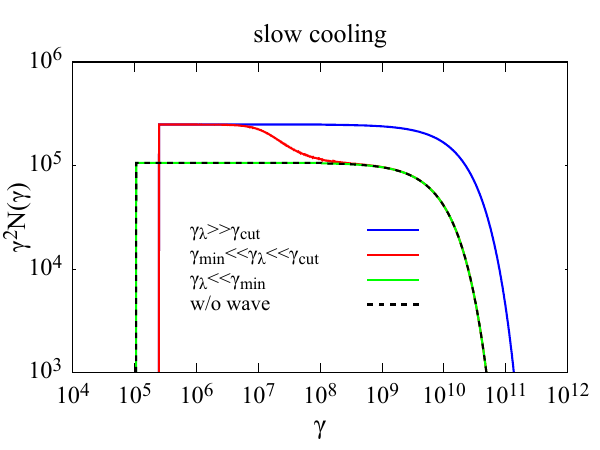}
  \includegraphics[width=9cm]{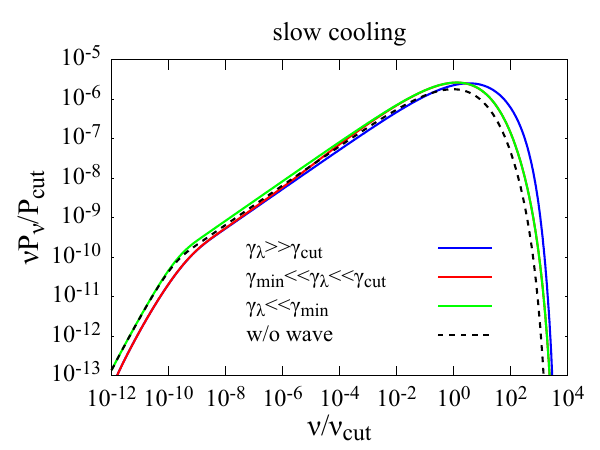}
  \caption{The electron energy distributions (upper) and the radiation spectra (lower) with a power-law ($p=2$) distribution for injection in the slow cooling case. The colored solid lines show the results for different $\gamma_\lambda$. The case without the wave is shown with the dashed black line.}
  \label{fig:disslow}
\end{figure}

As shown in the upper panel of Figure \ref{fig:disslow}, the electron energy distribution for $\gamma_\lambda \ll \gamma_{\rm min}$ (green) is not affected by the wave, and is almost the same as the case without the wave (dashed black).
On the contrary, for $\gamma_\lambda \gg \gamma_{\rm cut}$ (blue), all electrons gain energy by the wave, so that the distribution is shifted to higher energies.
In the case for $\gamma_{\rm min}<\gamma_\lambda<\gamma_{\rm cut}$ (red), the distribution shows a concave shape connecting the two cases of $\gamma_\lambda \ll \gamma_{\rm min}$ and 
$\gamma_\lambda \gg \gamma_{\rm cut}$ around $\gamma\simeq \gamma_\lambda=10^8$.

The lower panel of Figure \ref{fig:disslow} shows the resultant photon spectra.
The spectrum and frequency are normalized by
\begin{eqnarray}
P_{\mathrm{cut}} \equiv \frac{4}{3}\gamma_{\mathrm{cut}}^{2}c\sigma_{T}\frac{B^{2}}{8\pi},
\end{eqnarray}
and
\begin{eqnarray}
\nu_{\mathrm{cut}}\equiv\frac{3}{16}\gamma_{\mathrm{cut}}^{2}\frac{eB}{m_{\mathrm{e}}c},
\end{eqnarray}
respectively.
The boost of electron energies for $\gamma_\lambda \gg \gamma_{\rm cut}$ (blue) leads to a slightly higher peak energy in the photon spectrum. However, as the induced electric field suppresses the emissivity (see Figure \ref{fig:sperl}), the energy boost does not enhance the power in the low-frequency regime compared to the case without the wave (dashed black).
For $\gamma_\lambda \ll \gamma_{\rm min}$ (green), 
the emissivity is slightly higher compared to the case without the wave owing to the enhanced acceleration as discussed in \S \ref{sec:result1}.
For $\gamma_{\rm min}<\gamma_\lambda<\gamma_{\rm cut}$ (red), the radiation spectrum is curved slightly around
\begin{eqnarray}
\nu_{\lambda}\equiv\frac{3}{16}\gamma_{\lambda}^{2}\frac{eB}{m_{e}c},
\end{eqnarray}
connecting the blue line in low-frequency region and the green line in high-frequency region.
However, the modulation is not so prominent.

\subsection{Fast cooling}

\begin{figure}[h]
  \centering
  \includegraphics[width=9cm]{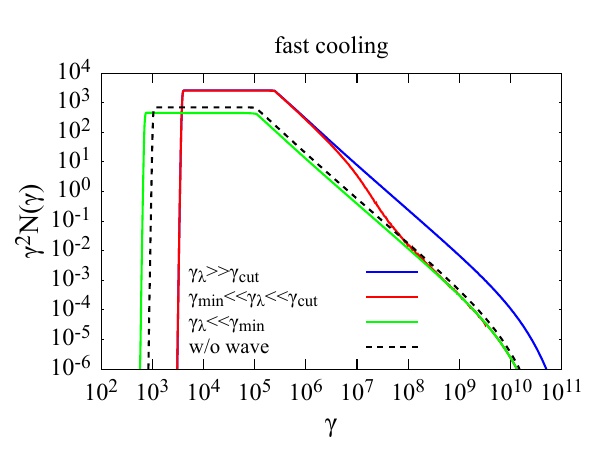}
  \includegraphics[width=9cm]{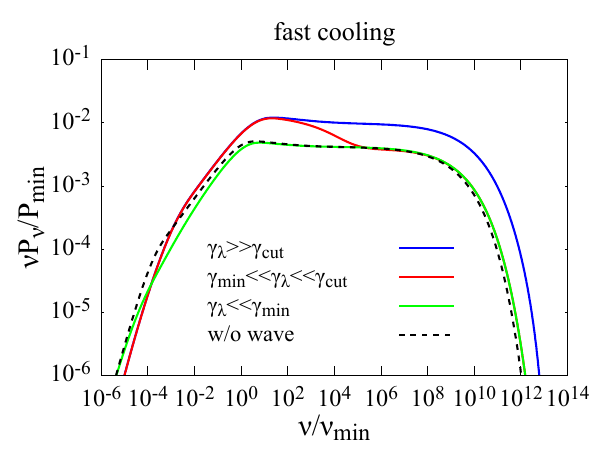}
  \caption{Same as Figure \ref{fig:disslow} but for the fast cooling case.}
  \label{fig:disfast}
\end{figure}

For the fast cooling ($\gamma_{\mathrm{c}}\ll \gamma_{\mathrm{min}}$) case, the electron energy distribution becomes softer than the injection spectrum, and a low-energy component below $\gamma_{\mathrm{min}}$ appears.
The upper panel of Figure \ref{fig:disfast} shows these properties. The low-energy sharp breaks correspond to $\gamma_{\mathrm{min}}$. 
The energy distribution is shifted to higher energies by the induced electric field for $\gamma_\lambda \gg \gamma_{\rm cut}$ (blue).
Note that, unlike the slow cooling case, the energy shift for this steeper spectrum leads to a much larger increase of $N(\gamma)$ for a given $\gamma$.
The low-energy cut-off at $\sim \gamma_{\mathrm{c}}$ is relatively high because of the cooling suppression.
The distributions for the other cases (green and red) present similar behaviors to those in the slow-cooling cases, though the differences in the cooling efficiency appear in the different low-energy cut-offs.

The normalization for the lower panel of Figure \ref{fig:disfast} is similar to the slow cooling case with
\begin{eqnarray}
P_{\mathrm{min}} \equiv \frac{4}{3}\gamma_{\mathrm{min}}^{2}c\sigma_{T}\frac{B^{2}}{8\pi},\\
\nu_{\mathrm{min}}\equiv\frac{3}{16}\gamma_{\mathrm{min}}^{2}\frac{eB}{m_{\mathrm{e}}c}.
\end{eqnarray}
The spectral power for $\gamma_\lambda \gg \gamma_{\rm cut}$ is high compared to the case without the wave, even though the radiation power is relatively suppressed by the induced electric field.
This is due to the large increase of $N(\gamma)$ by the energy shift for a steep energy distribution as we mentioned.
Even in the slow cooling case, this enhancement of emissivity can be expected for a steep injection spectrum with e.g. $p=3$.
In the case of $\gamma_\lambda \ll \gamma_{\rm c}$ shown with the green line, 
the radiation spectrum is shifted to a slightly higher frequency in the high-frequency region compared to the case without the wave.

In the case of $\gamma_{\rm min}<\gamma_\lambda<\gamma_{\rm cut}$ (red), the spectrum shows a characteristic bump structure due to the distorted electron spectrum.
The steeper spectrum above $\sim 10^2 \nu_{\rm min}$ extends to two orders of magnitude in $\nu$.
The flat component above $\sim 10^6 \nu_{\rm min}$ can be misunderstood as an extra synchrotron self-Compton component. In this misinterpretation, spectral modeling would lead to a much lower magnetic field than the actual value.

\section{Conclusions \& Discussion} \label{sec:sum}

Magnetically dominated outflow has been considered for pulsar winds, blazar jets, and gamma-ray bursts. Relativistic turbulence in highly magnetized plasma should induce electric fields with a large amplitude. While the standard synchrotron emission has been adopted for the model of emission from such objects, the large electric field induced in the outflows can affect the emission process from relativistic electrons.

As the first step to investigate the emission processes in relativistic turbulence, we have considered circularly polarized Alfv\'{e}n waves in this paper.
We have calculated the radiation spectrum by numerically following the electron trajectories in the waves, which can be analytically expressed.
We have shown the energy dependence of the emission power and spectrum from a single electron. 
For electrons whose Larmor radius is significantly smaller than the wavelength of the turbulence $r_{\rm L0}\ll\lambda$,
the motional electric field suppresses the emission power by a factor of $1/\Gamma_{E\times B}^{4}$, where $\Gamma_{E\times B}$ is the Lorentz factor for the $E\times B$ drift motion.
The value of $\Gamma_{E\times B}$ can be significantly large in relativistic turbulence.
However, in the case of the circularly polarized Alfv\'{e}n wave, $\Gamma_{E\times B}$ is limited to $\sqrt{2}$.
Note also that the average energy of electrons injected in the relativistic wave is boosted for $r_{\rm L0}\ll\lambda$.
For $r_{\rm L0}\gg\lambda$, the emission power and spectral peak frequency are slightly increased by the wave.

We have also demonstrated the emission spectra from electrons injected with a power-law energy distribution.
In the slow cooling case, though the complicated effects mentioned above are entangled, the resultant photon spectrum is not drastically modified. However, for the fast cooling case, the photon spectrum can be concave around $\nu_\lambda$, which is the typical photon energy emitted by electrons of $r_{\rm L0}\simeq\lambda$.

The resultant spectrum is similar to GRB photon spectra detected with {\it Fermi}-LAT \citep{2009ApJ...706L.138A,2010ApJ...712..558A,2010ApJ...716.1178A,2011ApJ...729..114A,2011ApJ...730..141Z,2014Sci...343...42A,2017A&A...606A..93Y}. The extra component detected in the GeV energy range has been interpreted as an SSC \citep{2010A&A...524A..92C,2010ApJ...720.1008C,2011arXiv1111.0127A,2012MNRAS.420..468P} or hadronic cascade component \citep{2009ApJ...705L.191A,2011arXiv1111.0127A,2011ApJ...739..103A}. If this extra component is due to relativistic waves in highly magnetized plasma, the estimate of the magnetic field and typical electron energy can be largely altered. In our demonstration, the caveat is that electrons are assumed to be isotropic and energetic as $r_{\rm L0}\gg\lambda$ at injection. If a wave comes just after the isotropic injection, our setup for the calculation is justified.

If the turbulence itself is responsible for particle injection/acceleration, the locally isotropic power-law injection may not be justified. As shown in \citet{2019ApJ...886..122C}, magnetic reconnection leads to an anisotropic particle injection. In such cases, depending on the induced anisotropy, the emission spectra would be modified \citep{2022ApJ...933...18G}.
In addition, multiple interactions with Alfv\'{e}n waves lead to reacceleration \citep[e.g.][]{2008ApJ...681.1725S,2019ApJ...877...71T}, which can also produce hard photon spectra in GRBs \citep{2009ApJ...705.1714A,2015MNRAS.454.2242A,2017ApJ...846L..28X}, blazars \citep{2011ApJ...740...64L,2014ApJ...780...64A,2018ApJ...861...31A}, and pulsar wind nebulae \citep{2017ApJ...841...78T,2023MNRAS.525.2750T}. Those combined effects lead to non-trivial shapes of photon spectra.

Though the existence of electrons with $r_{\rm L0}\gg\lambda$ is also non-trivial, there are several setups favorable for generating such high-energy electrons. For example, electrons are accelerated by the electric field due to the vacuum gap in the black hole magnetosphere \citep[e.g.][]{2007ApJ...671...85N}, then such electrons can be injected into a magnetically driven jet outside. In this case, the injection process and turbulence property in the emission site are independent. Another possible injection mechanism is magnetic reconnection and succeeding re-acceleration.
In this case, the injection scale of turbulence $\lambda$ is comparable to the scale of magnetic islands $l_{\rm MI}$.
The induced electric field at the reconnection site is roughly comparable to the background magnetic field.
The typical initial energy of accelerated electrons is $E_{\rm typ}\sim e E l_{\rm MI} \sim eB\lambda$.
The Larmor radii of electrons accelerated directly with the electric field in magnetic islands can be the same scale as the magnetic islands as $r_{\rm L0}=E_{\rm typ}/(eB)\sim \lambda$ 
\citep{2012ApJ...750..129B,2014ApJ...783L..21S,2019ApJ...886..122C}. These electrons can be further accelerated via stochastic non-gyro-resonant scattering off the turbulent fluctuations \citep[e.g. see ][]{2012PhRvL.108m5003H,2019ApJ...886..122C}. Such a process may cause an effective injection of electrons with $r_{\rm L0}\gg\lambda$. 

While we have focused on circularly polarized Alfv\'{e}n waves in this paper, our next step will be the investigation of emission in more general turbulence.
Especially for relativistic compressible waves, $\Gamma_{E\times B}$ can be significantly large. The induced large electric field may greatly modify the radiation spectrum.

\begin{acknowledgements}
We first appreciate the advice from the anonymous referee. We also thank Kosuke Nishiwaki, Yo Kusafuka, Tomoki Wada, Takumi Ohmura, Tomohisa Kawashima, Tomoya Kinugawa, Kyohei Kawaguchi for useful comments and discussion. 
R.G. acknowledges the support by the Forefront
Physics and Mathematics Program to Drive Transformation (FoPM).
This work is supported by the joint research program of the Institute for Cosmic Ray Research (ICRR), the University of Tokyo, and KAKENHI grant Nos. 22K03684, and 23H04899 (K.A.). 
Numerical computations were in part carried out on FUJITSU Server PRIMERGY CX2550 M5 at ICRR. 
\end{acknowledgements}

\begin{appendix}
\section{radiation of electrons injected isotropically into uniform fields} \label{sec:ebiso}
Electrons are injected isotropically with Lorentz factor $\gamma_{\mathrm{i}}$ into a uniform electric field $\bm{E}=E\bm{e_{y}}$ and a magnetic field $\bm{B}=B\bm{e_{z}}$.
The components of the  normalized velocity $\bm{\beta}=\frac{\bm{v}}{c}$ are $\beta_{x}=\beta_{0}\cos\theta$, $\beta_{y}=\beta_{0}\sin\theta\cos\phi$ and $\beta_{z}=\beta_{0}\sin\theta\sin\phi$.
The velocity of the $E\times B$ drift motion is $\bm{v}_{E\times B}=c\frac{\bm{E}\times\bm{B}}{B^{2}}=c\frac{E}{B}\bm{e_{x}}$.
In the drifting frame, the electron motion is spiral one around the magnetic field $\bm{B'}=\frac{B}{\Gamma_{E\times B}}\bm{e_{z}}$. Hereafter, the prime ' denotes quantities in the drifting frame.
The Lorentz factor is given by
\begin{eqnarray}
\label{eq:gs}
\gamma(\theta,\phi,t)=\Gamma_{E\times B}\gamma'(1+\beta_{E\times B}\beta'_{x}),
\end{eqnarray}
where $\beta_{E\times B} \equiv \frac{E}{B}$, $\Gamma_{E\times B} \equiv \frac{1}{\sqrt{1-\beta_{E\times B}^{2}}}$, $\beta'_{x}=\beta'\sin\alpha'\cos(\omega'_{B}t')$, $\alpha'$ is the pitch angle between the electron velocity and the magnetic field, and $\omega'_{B}=\frac{eB'}{\gamma'm_{e}c}$ is the gyro frequency.

We average $\gamma(\theta,\phi,t)$ over the period of $E\times B$ drift $T_{E\times B}=\Gamma_{E\times B}T'_{\mathrm{gyro}}$  where $T'_{\mathrm{gyro}}=\frac{2\pi}{\omega'_{B}}$ is the gyro period.
Then,
\begin{eqnarray}
\label{eq:gst0}
&&\langle\gamma(\theta,\phi)\rangle_{t}\equiv\frac{1}{T_{E\times B}}\int_{0}^{T_{E\times B}}\gamma(\theta,\phi,t)dt=\frac{1}{\Gamma_{E\times B}T'_{\mathrm{gyro}}}\int_{0}^{T'_{\mathrm{gyro}}}\Gamma^{2}_{E\times B}\gamma'(1+\beta_{E\times B}\beta'_{x}(t'))^{2}dt'\nonumber\\
&&=\Gamma_{E\times B}\gamma'\bigg\lparen1+\frac{1}{2}\beta_{E\times B}^{2}\beta'^{2}\sin^{2}\alpha'\bigg\rparen.
\end{eqnarray}

\begin{figure}[h]
  \centering
  \includegraphics[width=9cm]{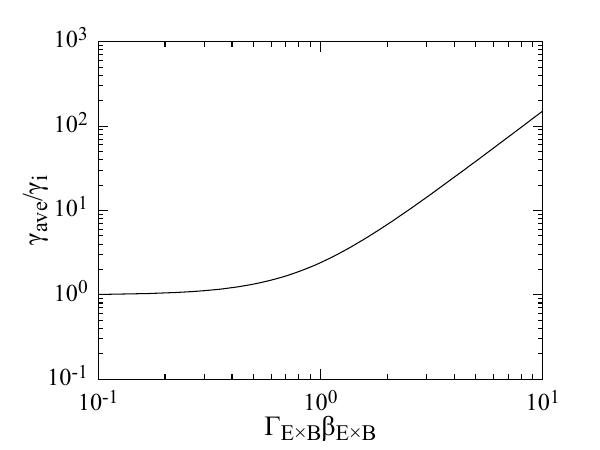}
  \caption{The average Lorentz factor $\gamma_{\mathrm{ave}}$ of electrons, which are isotropically injected with $\gamma_{\mathrm{i}}$. }
  \label{fig:gaveeb}
\end{figure}

Using $\gamma'=\Gamma_{E\times B}\gamma_{\mathrm{i}}(1-\beta_{E\times B}\beta_{0}\cos\theta)$, $\beta'^{2}\sin^{2}\alpha'=\beta'^{2}_{x}+\beta'^{2}_{y}$ and the Lorentz transformation of velocities $\beta'_{x}=(\beta_{x}-\beta_{E\times B})/(1-\beta_{E\times B}\beta_{x})$ and $\beta'_{y}=\beta_{y}/\Gamma_{E\times B}(1-\beta_{E\times B}\beta_{x})$,
we obtain
\begin{eqnarray}
\label{eq:gst}
&&\langle\gamma(\theta,\phi)\rangle_{t}=\Gamma_{E\times B}^{2}\gamma_{\mathrm{i}}f(\theta,\phi),
\end{eqnarray}
where
\begin{eqnarray}
\label{eq:f}
f(\theta,\phi)
=&&(1-\beta_{E\times B}\beta_{0}\cos\theta)\bigg\lbrack1+\frac{1}{2}\beta_{E\times B}^{2}\bigg\lbrace\frac{(\beta_{0}\cos\theta-\beta_{E\times B})^{2}}{(1-\beta_{E\times B}\beta_{0}\cos\theta)^{2}}+\frac{\beta_{0}^{2}\sin^{2}\theta\cos^{2}\phi}{\Gamma_{E\times B}^{2}(1-\beta_{E\times B}\beta_{0}\cos\theta)^{2}}\bigg\rbrace\bigg\rbrack.
\end{eqnarray}
The angle-averaged Lorentz factor is obtained as
\begin{eqnarray}
\label{eq:gaves}
\gamma_{\mathrm{ave}}&\equiv&\frac{1}{4\pi}\int_{0}^{2\pi}d\phi\int_{0}^{\pi}d\theta\sin\theta\langle\gamma(\theta,\phi)\rangle_{t}\nonumber\\
&=&\Gamma_{E\times B}^{2}\gamma_{\mathrm{i}}g(\beta_{E\times B}),
\end{eqnarray}
where
\begin{eqnarray}
\label{eq:g}
g(\beta_{E\times B})=\frac{3}{4}(1+\beta_{E\times B}^{2})+\frac{1-2\beta_{E\times B}^{2}+\beta_{0}^{2}\beta_{E\times B}^{2}}{8\Gamma_{E\times B}^{2}\beta_{0}\beta_{E\times B}}\ln\frac{1+\beta_{E\times B}\beta_{0}}{1-\beta_{E\times B}\beta_{0}}
\end{eqnarray}
For $\beta_{E\times B}=1/\sqrt{2}$ and $\beta_{0}\simeq 1$, $g(\beta_{E\times B})\simeq 1.20$.
As shown in Figure \ref{fig:gaveeb},  $\gamma_{\mathrm{ave}}$ is boosted from injected Lorentz factor $\gamma_{\mathrm{i}}$ by a factor of $\Gamma_{E\times B}^{2}$ for $\Gamma_{E\times B}\beta_{E\times B}\gg1$.

\begin{figure}[H]
  \centering
  \includegraphics[width=9cm]{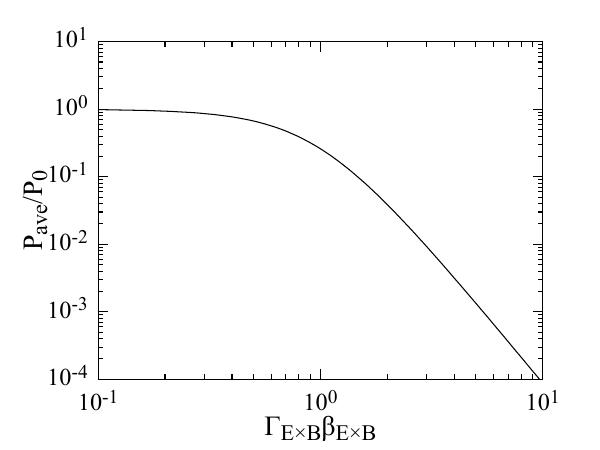}
  \caption{The average radiation power of electrons normalized by $P_0$.}
  \label{fig:powaveeb}
\end{figure}

Next, we estimate the radiation power.
In the drifting frame, the radiation power is constant.
As the radiation power is Lorentz invariant, the radiation power in the original frame is also constant and equal to the synchrotron power in the drifting frame.
The radiation power of an electron with injection angle $\theta,\phi$ is
\begin{eqnarray}
\label{eq:pows}
&&P(\theta,\phi)=2c\sigma_{T}\frac{B'^{2}}{8\pi}\gamma'^{2}\beta'^{2}\sin^{2}\alpha'=2c\sigma_{T}\frac{B^{2}}{8\pi}\gamma_{\mathrm{i}}^{2}\bigg\lbrace\bigg\lparen\beta_{0}\cos\theta-\beta_{E \times B}\bigg\rparen^{2}+\frac{\beta_{0}^{2}\sin^{2}\theta\cos^{2}\phi}{\Gamma_{E\times B}^{2}}\bigg\rbrace. 
\end{eqnarray}
The average power is calculated as
\begin{eqnarray}
\label{eq:powaves}
&&P_{\mathrm{ave}}\equiv\frac{1}{4\pi}\int_{0}^{2\pi}d\phi\int_{0}^{\pi}d\theta\sin\theta P(\theta,\phi)=2c\sigma_{T}\frac{B^{2}}{8\pi}\gamma_{\mathrm{i}}^{2}h(\beta_{E\times B}),
\end{eqnarray}
where
\begin{eqnarray}
\label{eq:h}
h(\beta_{E\times B})=\frac{1}{3}\beta_{0}^{2}+\beta_{E\times B}^{2}+\frac{1}{3}\frac{\beta_{0}^{2}}{\Gamma_{E\times B}^{2}}.
\end{eqnarray}
Using equations (\ref{eq:gaves}) and (\ref{eq:powaves}), we obtain
\begin{eqnarray}
\label{eq:powavegaves}
&&P_{\mathrm{ave}}=2c\sigma_{T}\frac{B^{2}}{8\pi}\gamma_{\mathrm{ave}}^{2}\frac{1}{\Gamma_{E\times B}^{4}}\frac{h(\beta_{E\times B})}{g(\beta_{E\times B})^{2}}.
\end{eqnarray}
The results are plotted in \textbf{Figure \ref{fig:powaveeb}}, which shows the suppression of the radiative cooling by a factor of $1/\Gamma_{E\times B}^{4}$ as discussed in the section \ref{sec:eb}.

\end{appendix}

\bibliography{ref}{}

\bibliographystyle{aasjournal}

\end{document}